\documentclass{article}
\usepackage[latin9]{inputenc}
\usepackage{amstext}

\makeatletter

\DeclareFontEncoding{LGR}{}{}
\DeclareRobustCommand{\greektext}{%
  \fontencoding{LGR}\selectfont\def\encodingdefault{LGR}}
\DeclareRobustCommand{\textgreek}[1]{\leavevmode{\greektext #1}}
\ProvideTextCommand{\~}{LGR}[1]{\char126#1}

\DeclareTextSymbolDefault{\textquotedbl}{T1}
\providecommand{\tabularnewline}{\\}

\newcommand{\lyxaddress}[1]{
	\par {\raggedright #1
	\vspace{1.4em}
	\noindent\par}
}

\makeatother

\begin{document}
\title{The solutions of the Yang-Baxter equation for the $(n+1)(2n+1)$ -vertex
models through a differential approach.}
\author{R. S. Vieira and A. Lima-Santos}
\maketitle

\lyxaddress{Universidade Federal de São Carlos, Departamento de Física, Caixa
Postal 676, CEP 13569-905, São Carlos, Brasil}
\begin{abstract}
The formal derivatives of the Yang-Baxter equation with respect to
its spectral parameters, evaluated at some fixed point of these parameters,
provide us with two systems of dierential equations. The derivatives
of the R matrix elements, however, can be regarded as independent
variables and eliminated from the systems, after which two systems
of polynomial equations are obtained in place. In general, these polynomial
systems have a non-zero Hilbert dimension, which means that not all
elements of the R matrix can be fixed through them. Nevertheless,
the remaining unknowns can be found by solving a few number of simple
differential equations that arise as consistency conditions of the
method. The branches of the solutions can also be easily analyzed
by this method, which ensures the uniqueness and generality of the
solutions. In this work we considered the Yang-Baxter equation for
the $(n+1)(2n+1)$-vertex models with a generalization based on the
$A_{n-1}$ symmetry. This differential approach allowed us to solve
the Yang-Baxter equation in a systematic way. . 
\end{abstract}
\vspace{1.2cm}
 
\begin{center}
\par\end{center}

\newpage{}

\section{Introduction}

A short life story: 

- \textquotedbl I was privileged to work with Ricardo Soares Vieira
or Ricardinho, as he was called by colleagues and friends. 

Ricardinho died prematurely on October 21, 2020, due to a post-operative
complication to remove stomach cancer. 

I followed his graduate studies in physics at the Federal University
of São Carlos and was his advisor for the master's \cite{vieira3}and
doctorate \cite{vieira2}. 

During this period it was possible to observe his interests in the
various areas of knowledge. 

I was often impressed by his ability to solve complex problems quickly\textquotedbl . 

I'm very sorry for his death! 

A few months before his death, Vieira came to me to propose me to
work with the Yang-Baxter equation.The problem was to consider the
possible vertex models, taking into account the structure of the R
matrices associated with the symmetries of the non-exceptional affine
Lie algebras. The new R matrix solution is recalculated by the Yang
Baxter equation using a differential approach. He had just solved
this problem for two-state models\cite{vieira 4} . Then we started
looking at the $A_{n-1}^{(1)}$models , in the certainty that they
are the simplest and due to their current interest \cite{bittl 1,bittl2}.
We have organized this paper as follows. In sections $2$ and $3$
we make a usual presentation of Yang-Baxter's equation and its corresponding
differential equations , respectively \cite{vieira 4}. In section
$4$, we presente the calculos for the $15$-vertex model. In section
5 we conside the case $n=3$, or $28$-vertex model and in section
$6$ we presente the general case. We conclude in the section $7.$

\section{The Yang-Baxter equation}

The Yang-Baxter equation (YBE) is one of the most important equations
of contemporary mathematical-physics. It originally emerged in two
different contexts of theoretical physics: in quantum field theory,
the YBE appeared as a sufficient condition for the many-body scattering
amplitudes to factor into the product of pairwise scattering amplitudes
{[}1,2,3{]}; in statistical mechanics it represented a sufficient
condition for the transfer matrix of a given statistical model to
commute for different values of the spectral parameters \cite{baxter1,baxter2}.
Since the pioneer works in quantum integrable systems -- see \cite{kulish1,jimbo1,kulish2}
for a historical background --, the YBE has become a cornerstone
in several fields of physics and mathematics: it is most known for
its fundamental role in the quantum inverse scattering method and
in the algebraic Bethe Ansatz \cite{jskly1,takh,skly2}, although
it also revealed to be important in the formulation of Hopf algebras
and quantum groups \cite{skly3,jimb1,drinfed1,drinfeld2,faddeev},
in knot theory \cite{turaev}, in quantum computation \cite{kauffman},
in AdS-CFT correspondence \cite{minahan,beisert} and, more recently,
in gauge theory \cite{witten,costello1,costello2}. The YBE can be
seen as a matrix relation defined in $End(V\otimes V\otimes V)$,
where $V$ is an $n$- dimensional complex vector space. In the most
general case, it reads:
\begin{equation}
Y\!B=R_{12}(u)R_{13}(u+v)R_{23}(v)-R_{23}(v)R_{13}(u+v)R_{12}(u)\label{YB}
\end{equation}
where the arguments u and v, called spectral parameters, have values
in C. The solution of the YBE is an R matrix defined in End (V \ensuremath{\otimes}
V ). The indexed matrices Rij appearing in (1) are defined in End
(V \ensuremath{\otimes} V \ensuremath{\otimes} V ) through the formulas
\begin{equation}
R_{12}=R\otimes I,\qquad R_{23}=I\otimes R,\qquad R_{13}=P_{23}R_{12}P_{23}
\end{equation}
where $I\in End(V$) is the identity matrix, $P\in End(V\otimes V)$
is the permutator matrix (defined by the relation $P(A\otimes B)P=B\otimes A$
for $\forall A,B\in End(V))$ and $P_{12}=P\otimes I,P_{23}=I\otimes P$. 

For each solution of the YBE, a given integrable system can be associated.
In fact, in statistical mechanics, the $R$ matrix represents the
Boltzmann weights of a given statistical model while, in quantum field
theory, the $R$ matrix is associated with factorizable scattering
amplitudes between relativistic particles. From the YBE we can prove
that systems described by an $R$ matrix possess infinitely many conserved
quantities in involution -- the Hamiltonian being one of them --,
the reason why they are called integrable \cite{korepin} We say that
a given solution $R(u$) of the YBE (1) is regular if $R(0)=P$. Regular
solutions of the YBE have several important properties {[}8{]}.

\section{The differential Yang-Baxter equation}

The YBE corresponds to a system of non-linear functional equations.
Several particular solutions of the YBE are known {[}6,7,8{]}. The
first solutions were found by a direct inspection of the functional
equations, which are in fact very simple because the $R$ matrix is
assumed to have many symmetries. Nevertheless, there are other more
advanced methods for solving the YBE: we can cite, for instance, the
Baxterization of braid relations \cite{jones}, the use of Lie algebras
and superalgebras \cite{jimb2,bazh1,bazh2}, the construction of Hopf
algebras and quantum groups \cite{jimb1,drinfed1,drinfeld2}, and
also techniques relying on algebraic geometry \cite{krich}, see also
\cite{pimen}. The methods mentioned above usually require that the
$R$ matrix presents one or more symmetries from the very start. From
a mathematical point of view, would be desirable to develop a method
that requires in principle as few as possible symmetries and, at the
same time, that is powerful enough in order to find and classify the
solutions of the YBE. This paper is concerned with the development
and extensive use of such a method, which is based on a differential
approach. 

From the quantum grup invariant representation for non-excepitional
affine Lie algebra $A_{n-1}^{(1)}$ \cite{jimb2},we consider the
following regeralization for a R-matrix solution of the Yang-Baxter
equation (YBE)

\begin{equation}
R(u)=\sum_{i=1}^{n+1}a_{ii}(u)\mathrm{e}_{ii}\otimes\mathrm{e}_{ii}+\sum_{i\neq j}^{n+1}b_{ij}(u)\mathrm{e}_{ii}\otimes\mathrm{e}_{jj}+\sum_{i\neq j}^{n+1}c_{ij}(u)\mathrm{\mathrm{e}}_{ij}\otimes\mathrm{e}_{ji}\label{eq:R-1}
\end{equation}
where $e_{ij}^{(n)}$ are the Weyl matrices acting for a $n+1$ dimensional
vector space $V$ at the site $n$ . The R-matrix elementes $a_{ii}(u)$,
$b_{ij}(u)$ and $c_{ij}(u)$ are fixed.

To be more precise, this method consists mainly of the following:
if we take the formal derivatives of the (\ref{YB}) with respect
to the spectral parameters $u$ and $v$ and then evaluate the derivatives
at some fixed point of those variables (say at zero), then we shall
get two systems of ordinary non-linear differential equations for
the elements of the R matrix. The derivatives of the $R$ matrix elements,
however, can be regarded as independent variables, so that, after
they are eliminated, two systems of polynomial equations are obtained
in place. Thus, these polynomial systems can be analyzed -- for instance,
through techniques of the computational algebraic geometry \cite{abel}
-- and eventually completely solved. It happens, however, that these
polynomial systems usually have a positive Hilbert dimension, which
means that the systems are satisfied even when some of the $R$ matrix
elements are still arbitrary. The remaining unknowns, nonetheless,
can be found by solving a few number of differential equations that
arise from the expressions for the derivatives we had eliminated before.
These auxiliary differential equations, therefore, can be thought
as consistency conditions of the method. For example, if we take the
formal derivative of (\ref{YB}) with respect to $v$ and then evaluate
the result at the point $v=0$, then we shall get the equation,

\begin{equation}
\textrm{\ensuremath{Y\!B_{v}}}=R_{12}(u){}_{13}(u)P_{23}+R_{12}(u)R_{13}(u)H_{23}-H_{23}R_{13}(u)R_{12}(u)-P_{23}R_{13}(u)R_{12}(u)\label{DR1}
\end{equation}
and from its derivative with respect $u$ at the point $u=0$, we
get 

\begin{equation}
Y\!B_{u}=H_{12}R_{13}(v)R_{23}(v)+P_{12}D_{13}(v)R_{23}(v)-R_{23}(v)D_{13}(v)P_{12}-R_{23}(v)R_{13}(v)H_{12}\label{DR2}
\end{equation}
where$P$ is the permutator matrix. $P=R(0)$ and $H=D(0)$. 

We highlight that $\mathcal{H}=PH$, where $\mathcal{H}$ is nothing
but the local Hamiltonian associated with the model -- see, for instance,
{[}6,8{]}. 

Therefore
\begin{equation}
D(u)=\frac{\partial R(u+v)}{\partial v}\mid_{v=0},\qquad D(v)=\frac{\partial R(u+v)}{\partial u}\mid_{u=0},\qquad P=R(0),\qquad H=D(0)
\end{equation}
Using the notation
\begin{equation}
\frac{da_{ii}(u)}{du}\mid_{u=0}=\alpha_{ii},\qquad\frac{db_{ij}(u)}{du}\mid_{u=0}=\beta_{ii}\qquad\mathrm{and}\qquad\frac{dc_{ij}(u)}{du}\mid_{u=0}=\mu_{ij}
\end{equation}
we can write

\begin{equation}
D(u)=\sum_{i=1}^{n+1}da_{ii}(u)\mathrm{e}_{ii}\otimes\mathrm{e}_{ii}+\sum_{i\neq j}^{n+1}db_{ij}(u)\mathrm{e}_{ii}\otimes\mathrm{e}_{jj}+\sum_{i\neq j}^{n+1}dc_{ij}(u)\mathrm{\mathrm{e}}_{ij}\otimes\mathrm{e}_{ji}\label{eq:D}
\end{equation}
and

\begin{equation}
H=\sum_{i=1}^{n+1}\alpha_{ii}\mathrm{e}_{ii}\otimes\mathrm{e}_{ii}+\sum_{i\neq j}^{n+1}\beta_{ij}\mathrm{e}_{ii}\otimes\mathrm{e}_{jj}+\sum_{i\neq j}^{n+1}\mu_{ij}\mathrm{\mathrm{e}}_{ij}\otimes\mathrm{e}_{ji}\label{D0}
\end{equation}
where $da,db,dc$ are the derivatives of $a,b,c$ in respect to u
or $v$.

The idea of transforming a functional equation into a differential
one goes back to the works of the Niels Henrik Abel, who solved several
functional equations in this way . Abel\textquoteright s method presents
many advantages when compared with other methods of solving functional
equations. For instance, it consists in a general method that can
be applied to a huge class of functional equations; it establishes
the generality and uniqueness of the solutions (which would be difficult,
if not impossible, to establish in other ways) by reducing the problem
to the theory of differential equations and so on -- see {[}32{]}
for more. Notice moreover that although Abel\textquoteright s method
requires the solutions to be differentiable (there can be non-differentiable
solutions of some functional equations), this restriction is not a
problem when dealing with the YBE, as its solutions are always assumed
to be differentiable because of the connection between the R matrix
and the corresponding local Hamiltonian. Concerning the theory of
integrable systems, the differential method is perhaps most known
in connection with boundary YBE \cite{skly4,nepo,virira1}.

Now we can look for the matrices (\ref{eq:R-1}) what are solutions
of (\ref{YB}). First, let's explain the calculations for the deformed
$A_{1}^{(1)},$ or $15$ -vertex model

\section{The $15$-vertex models}

The correspondig matrices are

\begin{equation}
R(u)=\left(\begin{array}{ccccccccc}
a_{11} & 0 & 0 & 0 & 0 & 0 & 0 & 0 & 0\\
0 & b_{12} & 0 & c_{12} & 0 & 0 & 0 & 0 & 0\\
0 & 0 & b_{13} & 0 & 0 & 0 & c_{13} & 0 & 0\\
0 & c_{21} & 0 & b_{21} & 0 & 0 & 0 & 0 & 0\\
0 & 0 & 0 & 0 & a_{22} & 0 & 0 & 0 & 0\\
0 & 0 & 0 & 0 & 0 & b_{23} & 0 & c_{23} & 0\\
0 & 0 & c_{31} & 0 & 0 & 0 & b_{31} & 0 & 0\\
0 & 0 & 0 & 0 & 0 & c_{32} & 0 & b_{32} & 0\\
0 & 0 & 0 & 0 & 0 & 0 & 0 & 0 & a_{33}
\end{array}\right)\label{eq:R}
\end{equation}
\begin{equation}
D(u)=\left(\begin{array}{ccccccccc}
da_{11} & 0 & 0 & 0 & 0 & 0 & 0 & 0 & 0\\
0 & db_{12} & 0 & dc_{12} & 0 & 0 & 0 & 0 & 0\\
0 & 0 & db_{13} & 0 & 0 & 0 & dc_{13} & 0 & 0\\
0 & dc_{21} & 0 & db_{21} & 0 & 0 & 0 & 0 & 0\\
0 & 0 & 0 & 0 & da_{22} & 0 & 0 & 0 & 0\\
0 & 0 & 0 & 0 & 0 & db_{23} & 0 & dc_{23} & 0\\
0 & 0 & dc_{31} & 0 & 0 & 0 & db_{31} & 0 & 0\\
0 & 0 & 0 & 0 & 0 & dc_{32} & 0 & db_{32} & 0\\
0 & 0 & 0 & 0 & 0 & 0 & 0 & 0 & da_{33}
\end{array}\right)
\end{equation}
\begin{equation}
H=\left(\begin{array}{ccccccccc}
\alpha11 & 0 & 0 & 0 & 0 & 0 & 0 & 0 & 0\\
0 & \beta_{12} & 0 & \mu_{12} & 0 & 0 & 0 & 0 & 0\\
0 & 0 & \beta_{13} & 0 & 0 & 0 & \mu_{13} & 0 & 0\\
0 & \mu_{21} & 0 & \beta_{21} & 0 & 0 & 0 & 0 & 0\\
0 & 0 & 0 & 0 & \alpha_{22} & 0 & 0 & 0 & 0\\
0 & 0 & 0 & 0 & 0 & \beta_{23} & 0 & \mu_{23} & 0\\
0 & 0 & \mu_{31} & 0 & 0 & 0 & \beta_{31} & 0 & 0\\
0 & 0 & 0 & 0 & 0 & \mu_{32} & 0 & \beta_{32} & 0\\
0 & 0 & 0 & 0 & 0 & 0 & 0 & 0 & \alpha_{33}
\end{array}\right)
\end{equation}
Note the entries of $R(u)$ and $D(u)$ are functions of $u$ and
the conditions $a_{ii}(0)=1,b_{ij}(0)=0$ and $c_{ij}(0)=1$ , define
the matrix $\text{P}$, $P=R(0)$.

For this model we have three $27$ by$27$ matrices equations $YB=0,YB_{u}=0$
and $YB_{v}=0$. Looking at their diagonals $Y\!B[i.i]=0$, we can
find several equations containing only the $c_{ij}(u)$ amplitudes.
For their derivatives, $YB_{u}[i,i]=0$, we find

\begin{equation}
c_{ij}(u)c_{ji}(u)\left(\mu_{ij}-\mu_{ji}\right)+c_{ij}(u)dc_{ji}(u)-dc_{ij}(u)c_{ji}(u)=0,\quad i\neq j=\{1,2,3\}
\end{equation}
With the regular condions, the solutions are

\begin{equation}
c_{ij}(u)=\exp(\mu_{ij}u)\label{eq:C}
\end{equation}
After replace the $c_{ij}(u)$in all remained equations , the conjugated
equations $Y\!B[i,28-i]=0,Y\!B_{u}[i,28-i]=0$ and $Y\!B_{v}[i,28-i]=0$
are solved by the following relations

\begin{equation}
b_{23}(u)=\frac{\beta_{23}}{\beta_{21}}b_{21}(u),\qquad b_{32}(u)=\frac{\beta_{32}}{\beta_{12}}b_{12}(u)
\end{equation}
 with the constraint $\beta_{23}\beta_{32}=\beta_{12}\beta_{21}$
. From other equations find $b_{13}(u),b_{31}(u)$and its derivatives

\begin{equation}
b_{13}(u)=\frac{\beta_{13}}{\beta_{12}}b_{12}(u),\qquad b_{31}(u)=\frac{\beta_{31}}{\beta_{21}}b_{21}(u),
\end{equation}
with a second constraind $\beta_{13}\beta_{31}=\beta_{12}\beta_{21}.$

We notice that these relations can be write in a more compact form

\begin{equation}
b_{ij}(u)=\beta_{ij}K(u),\qquad\beta_{ji}=\frac{\beta_{12}\beta_{21}}{\beta_{ij}},\quad i\neq j=\{1,2,3\}\label{eq:bfunc}
\end{equation}
where $K(u)$ is an srbitrary funcion, to be fixed. 

Using (\ref{eq:bfunc}), all remaind equations contain only the $a_{ii}(u)$
and $K(u)$, as well as its derivatives. In particular, the differential
equation $YB_{u}[6,20]=0$ allow us to find the function $K(u)$:

\begin{equation}
-\frac{d}{du}K(u)+\left(\mu_{13}+\mu_{21}-\mu_{23}\right)K(u)+\exp((\mu_{12}+\mu_{23}-\mu_{13})u)=0
\end{equation}
For regular solutions we have

\begin{equation}
K(u)=\frac{\exp((\mu_{12}-\mu_{13}+\mu_{23})u)-\exp((\mu_{21}+\mu_{13}-\mu_{23})u)}{(\mu_{12}-2\mu_{13}-\mu_{21}+2\mu_{23})}\label{K}
\end{equation}
Substituing in the Yang-Baxter equations we have to fix two $\mu_{ij}$

\begin{equation}
\mu_{31}=\mu_{21}+\mu_{12}-\mu_{13}.\qquad\mu_{32}=\mu_{21}+\mu_{12}-\mu_{23}
\end{equation}
Before the computation of the $a_{ii}(u),i=1,2,3$ ,we can still simplify
the notation by defining two parameters

\begin{equation}
\kappa_{1}=\mu_{13}+\mu_{21}-\mu_{23},\qquad\kappa_{2}=\mu_{12}-\mu_{13}+\mu_{23}\label{K1}
\end{equation}
It follows 

\begin{equation}
K(u)=\frac{e^{\kappa_{1}u}-e^{\kappa_{2}u}}{\kappa_{1}-\kappa_{2}},\qquad\mu_{3j}=\kappa_{1}+\kappa_{2}-\mu_{j3},\qquad j=1,2\label{K2}
\end{equation}
and from $YB_{u}[2,10]=0$ , we get the the relation between the parameters
$\alpha_{11,}\beta_{ij}$ and $\mu_{ij}$:

\begin{equation}
\beta_{21}\beta_{12}=(\kappa_{1}-\alpha_{11})(\kappa_{2}-\alpha_{11})
\end{equation}
After we find the $a_{ii}(u)$ terms we get the following recurrence:

\begin{equation}
a_{ii}(u)=a_{11}(u)+(\alpha_{ii}-\alpha_{11})K(u),\qquad i=2,3
\end{equation}
with

\begin{equation}
a_{11}(u)=\frac{(\kappa_{1}-\alpha_{11})\mathrm{e^{\kappa_{2}u}-(\kappa_{2}-\alpha_{11})\mathrm{e^{\kappa_{1}u}}}}{\kappa_{1}-\kappa_{2}}
\end{equation}
Finally, we have a system of two equations whose solutions will determine
the $R$ matrices of the model.

\begin{equation}
(\alpha_{11}-\alpha_{kk})(\kappa_{1}+\kappa_{2}-\alpha_{11}-\alpha_{kk})=0,\qquad k=\{2,3\}\label{eq:sol}
\end{equation}
n this case we get four solutions:

\subsection{Solution 1: $\alpha_{22}=\alpha_{11},\qquad\alpha_{33}=\alpha_{11}$}

For this solution we have (\ref{eq:R}) with the following entries

\begin{equation}
c_{ij}(u)=\exp(\mu_{ij}(u)),\qquad b_{ij}(u)=\beta_{ij}K(u),\qquad i\neq j=\{1,2,3\}
\end{equation}
and

\begin{equation}
a_{33}(u)=a_{22}(u)=a_{11}(u)=\frac{(\kappa_{1}-\alpha_{11})\exp(\kappa_{2}u)-(\kappa_{2}-\alpha_{11})\exp(\kappa_{1}u)}{\kappa_{1}-\kappa_{2}}
\end{equation}
where
\begin{equation}
\kappa_{1}=\mu_{13}+\mu_{21}-\mu_{23},\qquad\kappa_{2}=-\mu_{13}+\mu_{12}+\mu_{23},\qquad K(u)=\frac{\exp(\kappa_{1}u)-\exp(\kappa_{2}u)}{\kappa_{1}-\kappa_{2}}
\end{equation}
and the fixed parameters

\begin{equation}
\mu_{31}=\mu_{12}+\mu_{21}-\mu_{13},\qquad\mu_{32}=\mu_{12}+\mu_{21}-\mu_{32}
\end{equation}
\begin{equation}
\beta_{ji}=\frac{\beta_{12}\beta_{21}}{\beta_{ij}},\quad i<j=\{1,2,3\},\qquad\beta_{12}\beta_{21}=(\kappa_{1}-\alpha_{11})(\kappa_{2}-\alpha_{11})
\end{equation}
For a particular choice of parameters

\begin{equation}
\mu_{ij}=\eta,\qquad\mu_{ji}=0,\quad(i<j),\qquad\beta_{ij}=\xi,\quad i\neq j=\{1,2,3\}
\end{equation}
with 
\begin{equation}
\eta=\frac{\xi^{2}-\alpha_{11}^{2}}{\alpha_{11}}
\end{equation}
we get the quantum group invariant solution of \cite{jimb2}

\subsection{Solution 2: $\alpha_{22}=\alpha_{11},\alpha_{33}=\kappa_{1}+\kappa_{2}-\alpha_{11}$}

In this case we get

\begin{equation}
a_{22}(u)=a_{11}(u)=\frac{(\kappa_{1}-\alpha_{11})\exp(\kappa_{2}u)-(\kappa_{2}-\alpha_{11})\exp(\kappa_{1}u)}{\kappa_{1}-\kappa_{2}}
\end{equation}
and
\begin{equation}
a_{33}(u)=\frac{(\kappa_{1}-\alpha_{11})\exp(\kappa_{1}u)-(\kappa_{2}-\alpha_{11})\exp(\kappa_{2}u)}{\kappa_{1}-\kappa_{2}}
\end{equation}

\subsection{Solution 3 : $\alpha_{22}=\kappa_{1}+\kappa_{2}-\alpha_{11},\qquad\alpha_{33}=\alpha_{11}$}

Here we get

\begin{equation}
a_{33}(u)=a_{11}(u)=\frac{(\kappa_{1}-\alpha_{11})\exp(\kappa_{2}u)-(\kappa_{2}-\alpha_{11})\exp(\kappa_{1}u)}{\kappa_{1}-\kappa_{2}}
\end{equation}
and
\begin{equation}
a_{22}(u)=\frac{(\kappa_{1}-\alpha_{11})\exp(\kappa_{1}u)+(\kappa_{2}-\alpha_{11})\exp(\kappa_{2}u)}{\kappa_{2}-\kappa_{1}}
\end{equation}

\subsection{Solution 4: $\alpha_{22}=\kappa_{1}+\kappa_{2}-\alpha_{11},\qquad\alpha_{33}=\kappa_{1}+\kappa_{2}-\alpha_{11}$}

In this case
\begin{equation}
a_{11}(u)=\frac{(\kappa_{1}-\alpha_{11})\exp(\kappa_{2}u)-(\kappa_{2}-\alpha_{11})\exp(\kappa_{1}u)}{\kappa_{1}-\kappa_{2}}
\end{equation}
and
\begin{equation}
a_{33}(u)=a_{22}(u)=\frac{(\kappa_{1}-\alpha_{11})\exp(\kappa_{2}u)-(\kappa_{2}-\alpha_{11})\exp(\kappa_{1}u)}{\kappa_{1}-\kappa_{2}}
\end{equation}

Here we remember that $\kappa_{1}+\kappa_{2}=\mu_{12}+\mu_{21}$ and
from (\ref{eq:sol}) that the $a_{ii}(u)$have only two values

\begin{equation}
a_{ii}(u)=A(u)=\frac{(\kappa_{1}-\alpha_{11})\exp(\kappa_{2}u)-(\kappa_{2}-\alpha_{11})\exp(\kappa_{1}u)}{\kappa_{1}-\kappa_{2}}
\end{equation}
when $\alpha_{ii}=\alpha_{11}$ , and
\begin{equation}
a_{jj}(u)=B(u)=\frac{(\kappa_{1}-\alpha_{11})\exp(\kappa_{1}u)-(\kappa_{2}-\alpha_{11})\exp(\kappa_{2}u)}{\kappa_{1}-\kappa_{2}}
\end{equation}
when $\alpha_{jj}=\kappa_{1}+\kappa_{2}-\alpha_{11}$. 

Therefore we have 

\begin{tabular}{|c|c|c|c|}
\hline 
 & $a_{11}(u)$ & $a_{22}(u)$ & $a_{33}(u)$\tabularnewline
\hline 
\hline 
solution 1 & $A(u)$ & $A(u$) & $A(u)$\tabularnewline
\hline 
solution 2 & $A(u)$ & $A(u)$ & $B(u)$\tabularnewline
\hline 
solution 3 & $A(u)$ & $B(u)$ & $A(u)$\tabularnewline
\hline 
solution 4 & $A(u)$ & $B(u)$ & $B(u)$\tabularnewline
\hline 
\end{tabular}

The solutions 2 and 3 are equivalent. It means That there are $3$
different solutions.

For these R matrix we start with $15$ parameters, the derivates of
matrix elentes at the point $u=0$. We fix, \{$\alpha_{22},\alpha_{33},\beta_{21},\beta_{31},\beta_{32},\mu_{31},\mu_{32}\}$.
Therefore these R matrices have $8$ - free parameters. 

\section{The $28$$\:$vertex - model }

In the $A_{2}^{(1)}$ model we have $28$-vertex model and its Yang-Baxter
solutions are obtained following the procedures used in the $15$
-vertex model. The $16$ by 1$6$ matrices$R,D,H$ are given (\ref{eq:R-1}),
(\ref{eq:D}) and (\ref{D0}) , with $n=3$, respectively. 

The diagonal matrix entries equations 
\begin{equation}
YB_{u}[i,i]=0,\qquad and\qquad YB_{v}[i,i]=0
\end{equation}
 are verified by the amplitudes 
\begin{equation}
c_{ij}(u)=\exp(\mu_{ij}u),\qquad i\neq j=\{1,2.3,4\}
\end{equation}
 where $\mu_{ij}$ are arbitary parameters to be fixed.

The $b_{ij}(u)$ vertices were computed as we did in the case of the
15 vertex model. Its form is 
\begin{equation}
b_{ij}(u)=\beta_{ij}K(u),\qquad i\neq j=\{1,2,3,4\}
\end{equation}
The parameters $\beta_{ij}$satisfy the relation 
\begin{equation}
\beta_{ji}=\frac{\beta_{12}\beta_{21}}{\beta_{ij}}=\frac{(\kappa_{1}-\alpha_{11})(\kappa_{2}-\alpha_{11})}{\beta_{ij}},\qquad i<j=\{1,2,3,4\}
\end{equation}
Now we can get the relations between the parameters $\mu_{ij}$from
the equations $\mathbf{YB_{v}[i,j]=0}$ and $\mathbf{YB_{u}[i,j]=0:}$
\begin{equation}
\mu_{3i}=\mu_{12}+\mu_{21}-\mu_{i3},\quad i<3\quad and\quad\mu_{4i}=\mu_{12}+\mu_{21}-\mu_{i4}\quad i<4
\end{equation}

Replacing the expressions in (\ref{eq:R-1}) and its derivatives (\ref{DR1})
and (\ref{DR2}), we find, for instance, from $YB_{v}[46,55]=0$ the
same function $K(u)$ \ref{K2}
\begin{equation}
K(u)=\frac{e^{\kappa_{1}u}-e^{\kappa_{2}u}}{\kappa_{1}-\kappa_{2}},
\end{equation}
where $\kappa_{1}=\mu_{13}+\mu_{21}-\mu_{23},\qquad\kappa_{2}=\mu_{12}-\mu_{13}+\mu_{23}$
. The $a_{ii}(u)$ functions still satisfy the recurrence
\begin{equation}
a_{kk}(u)=a_{11}(u)+(\alpha_{kk}-\alpha_{11})K(u),\qquad k=2,3,4
\end{equation}
 with
\begin{equation}
a_{11}(u)=\frac{(\kappa_{1}-\alpha_{11})\mathrm{e^{\kappa_{2}u}-(\kappa_{2}-\alpha_{11})\mathrm{e^{\kappa_{1}u}}}}{\kappa_{1}-\kappa_{2}}
\end{equation}
As we can see, the results are the same of the $15$ -vertex model,
but including the index $n+1=4.$

Unlike the previous case, the parameters $\mu_{ji},\qquad i<j$ are
not sufficient to fix the solutions. looking at the equations $Y\!\!B_{v}[i,j]=0$
we see that many of them are type $(\alpha_{kk}-\alpha_{11})(\alpha_{11}+\kappa_{1}+\kappa_{2}-\alpha_{kk})G_{ij}(u)=0$
and those that are not, will stay in that shape after we get some
parameters $\mu_{ij}$ but now with $i<j.$ These calculations are
very annoying, but the results are simple. We found two possibilities:
\begin{equation}
\mu_{ij}=\kappa_{1}-\mu_{1i}+\mu_{1j}\quad and\qquad\mu_{ij}=\kappa_{2}-\mu_{1i}+\mu_{1j}\label{const}
\end{equation}
For this model we have to fix two parameters $\mu_{24}$ and $\mu_{34}$.
After this we have three equations
\begin{equation}
\alpha_{kk}-\alpha_{11})(\alpha_{11}+\kappa_{1}+\kappa_{2}-\alpha_{kk})=0
\end{equation}
 and eight solutions for each set of fixed parameters

\begin{tabular}{|c|c|c|c|c|}
\hline 
 & $a_{11}(u)$ & $a_{22}(u)$ & $a_{33}(u)$ & $a_{44}(u)$\tabularnewline
\hline 
\hline 
sol 1 & $A(u)$ & $A(u)$ & $A(u)$ & $A(u\text{) }$\tabularnewline
\hline 
sol 2 & $A(u)$ & $A(u)$ & $A(u)$ & $B(u)$\tabularnewline
\hline 
sol 3 & $A(u)$ & $A(u)$ & $B(u)$ & $A(u)$\tabularnewline
\hline 
sol 4 & $A(u)$ & $A(u)$ & $B(u)$ & $B(u)$\tabularnewline
\hline 
sol 5 & $A(u)$ & $B(u)$ & $A(u)$ & $A(u)$\tabularnewline
\hline 
sol 6 & $A(u)$ & $B(u)$ & $A(u)$ & $B(u)$\tabularnewline
\hline 
sol 7 & $A(u)$ & $B(u)$ & $B(u)$ & $A(u)$\tabularnewline
\hline 
sol 8 & $A(u)$ & $B(u)$ & $B(u)$ & $B(u)$\tabularnewline
\hline 
\end{tabular}

Taking into account the equivalences for the solutions with the same
number of A(u) and B(u), we have $4$ different solutions.

Remember that
\begin{equation}
A(u)=\frac{(\kappa_{1}-\alpha_{11})\exp(\kappa_{2}u)-(\kappa_{2}-\alpha_{11})\exp(\kappa_{1}u)}{\kappa_{1}-\kappa_{2}}
\end{equation}
for $\alpha_{kk}=\alpha_{11}$ and
\begin{equation}
B(u)=\frac{(\kappa_{1}-\alpha_{11})\exp(\kappa_{1}u)-(\kappa_{2}-\alpha_{11})\exp(\kappa_{2}u)}{\kappa_{1}-\kappa_{2}}
\end{equation}
for $\alpha_{kk}=\alpha_{11}+\kappa_{1}+\kappa_{2}.$

Note that the second set of equations due to (\ref{const})has only
two amplitudes $c_{ij}(u).$ different,$c_{24}$ and $c_{34}$. Note
that we have fixed $7$ $\mu_{ij}$, $6$ $\beta_{ij}$ and $3$ $\alpha_{ii}$
. Therefore our $R$ matrices solutions have $12$ free-parameters.

Now we know how to generalize the results:

\section{The $(n+1)(2n+1)$ -vertex models}

For each value of $n$\textgreater 1, the R matrix has $n+1$ diagonal
entries $a_{ii}(u)$, that are determined by recurrence relative to
$a_{11}(u)$

\begin{equation}
a_{kk}(u)=a_{11}(u)+(\alpha_{kk}-\alpha_{11})K(u),\qquad k=2,...,n+1
\end{equation}
where 
\begin{equation}
\alpha_{jj}=\left(\frac{d}{du}a_{jj}(u)\right)_{u=0}\quad and\quad K(u)=\frac{e^{\kappa_{1}u}-e^{\kappa_{2}u}}{-\kappa_{2}+\kappa_{1}}
\end{equation}
where $\kappa_{1}=\mu_{13}+\mu_{21}-\mu_{23}$ and $\kappa_{2}=-\mu_{13}+\mu_{12}+\mu_{23}.$

The remaining $n(n+1)$diagonal entries 
\begin{equation}
b_{ij}(u)=\beta_{ij}K(u),\qquad i\neq j=\{1,...,n+1\}
\end{equation}
where 
\begin{equation}
\beta_{ij}=\left(\frac{d}{du}b_{ij}(u)\right)_{u=0}
\end{equation}
with the constraints
\begin{equation}
\beta_{ji}\beta_{ij}=\beta_{21}\beta_{12}=(\kappa_{1}-\alpha_{11})(\kappa_{2}-\alpha_{11})
\end{equation}
 The number of fixed parameters $\beta_{ij}$ is $n-1)(n+2)/2+1$.

The $n(n+1)$ off-diagonal matrix elements
\begin{equation}
c_{ij}(u)=e^{\mu_{ij}u}
\end{equation}
All $\mu_{ij}$ of the $c_{ij}(u)$ below od main diagonal are fixed
by the relation
\begin{equation}
\mu_{ji}=\kappa_{1}+\kappa_{2}-\mu_{ij},\qquad j>i
\end{equation}
The number is $(n-1)(n+2)/2$and some $\mu_{ij}$ above the main diagonal
can be fixed by two relation
\begin{equation}
\mu_{ij}=\kappa_{1}-\mu_{1i}+\mu_{1j}\quad and\quad\mu_{ij}=\kappa_{2}-\mu_{1i}+\mu_{1j},\quad j<i
\end{equation}
The number is $(n+1)(n-2)/2$. It means that we have two sets of solutions.
With these relations the Yang-Baxter equation and its derivatives
are solved by two sets of $2^{n}$ solutions of the following $n$
equations
\begin{equation}
(\alpha_{kk}-\alpha_{11})(\alpha_{kk}+\alpha_{11}-\kappa_{1}-\kappa_{2})=0
\end{equation}

Therefore we have two differentes values for the $a_{kk}(u)$ and
$n$ parameters $\alpha_{ii}$ are fixed. It mean that our R matrix
solutions have $n(n+3)/2+3$ free parameters

\begin{equation}
a_{kk}(u)=A(u)=\frac{(\kappa_{1}-\alpha_{11})\exp(\kappa_{2}u)-(\kappa_{2}-\alpha_{11})\exp(\kappa_{1}u)}{\kappa_{1}-\kappa_{2}}
\end{equation}
when $\alpha_{ii}=\alpha_{11}$ and
\begin{equation}
a_{kk}(u)=B(u)=\frac{(\kappa_{1}-\alpha_{11})\exp(\kappa_{1}u)-(\kappa_{2}-\alpha_{11})\exp(\kappa_{2}u)}{\kappa_{1}-\kappa_{2}}
\end{equation}
when $\alpha_{kk}=\kappa_{1}+\kappa_{2}-\alpha_{11}$. 

Using the identity
\begin{equation}
2^{n}=\sum_{k=0}^{n+1}\left(\begin{array}{c}
n\\
k
\end{array}\right)=\sum_{k=0}^{n+1}\frac{n!}{k!(n-k)!}
\end{equation}
we can identify $n+1$ different solutions

\section{Conclusion}

From our calculus for $n>2$, we have find two sets of $n+1$ $R$
matrix with $\frac{n(n+3)}{2}+3$ free parameters as solution of the
Yang-Baxter for the $(n+1)(2n+1)$vertex models. For a particular
choice of the parameters, the solutions with $n+1$ $A(u)$ function,
we recover the $R$ of the affine Lie algebra $A_{n-1}^{(1)}$. 

Several particular solutions of the Yang-Baxter equation associated
with The $15$ --vertex models are known. We can cite, for example,
the fifteen-vertex R matrices of Cherednik \cite{chere}, Babelon
\cite{babe}, Chudnovsky \& Chudnovsky \cite{chude} and Perk \& Schultz
\cite{perk1,perk2} and \cite{perk3} (these solutions hold for higher
vertex models as well). These R matrices contain fewer parameters
than the solutions we found, so that they can be thought of as reductions
of a more general solution.

We believe that the results presented here are original. 

Perhaps the calculation of the reflection matrices for these R matrices
can also be interesting, as well as their Bethe's ansatz.

\section*{Acknowledgement}

A posthumous thanks to Ricardinho. We thank R. A. Pimenta to correct
several misprints. This work was supported in part by Conselho Nacional
de Desenvolvimento-CNPq-Brasil.

\end{document}